\documentclass[12pt]{article}
\usepackage{graphicx}

\begin{document}

\title{Nuclear in-medium effects on $\eta$ dynamics in proton-nucleus collisions}

\author{Jie Chen$^{1,2}$, Zhao-Qing Feng$^{1,3}$  \footnote{Corresponding author. Tel. +86 931 4969152. \newline \emph{E-mail address:} fengzhq@impcas.ac.cn (Z.-Q. Feng)}, Jian-Song Wang$^{1}$}

\date{}
\maketitle

\begin{center}
$^{1}${\small \emph{Institute of Modern Physics, Chinese Academy of Sciences, Lanzhou 730000, People's Republic of China }}           \\
$^{2}${\small \emph{University of Chinese Academy of Sciences, Beijing 100190, People's Republic of China }}                          \\
$^{3}${\small \emph{Kavli Institute for Theoretical Physics, Chinese Academy of Sciences, Beijing 100190, People's Republic of China }}
\end{center}

\textbf{Abstract}
\par
The dynamics of $\eta$ meson produced in proton-induced nuclear reactions via the decay of N$^{\ast}(1535)$ has been investigated within the Lanzhou quantum molecular dynamics transport model (LQMD). The in-medium modifications of the $\eta$ production in dense nuclear matter are included in the model, in which an attractive $\eta$-nucleon potential is implemented. The impact of the $\eta$ optical potential on the $\eta$ dynamics is investigated. It is found that the attractive potential leads to the reduction of high-momentum (kinetic energy) production from the spectra of momentum distributions and inclusive cross sections and increasing the reabsorption process by surrounding nucleons.

\bigskip

\section{Introduction}

The in-medium properties of hadrons is one of topical issues in nuclear physics, in particular related to the chiral symmetry restoration, phase-transition from quark-gluon plasma to hadrons, nuclear equation of state (EoS), structure of neutron star etc \cite{Gi95,Fr07,St05}. Theoretically, because of the asymptotic freedom of Quantum Chromodynamics (QCD), it is hard to directly obtain the in-medium properties of hadrons from QCD. Therefore, many effective field models have been used. Heavy-ion collisions in terrestrial laboratory provide the way to study the in-medium properties of hadrons in dense nuclear matter and to extract the high-density behavior of the nuclear symmetry energy (isospin asymmetric part of EoS) \cite{Li08}. To understand the experimental data from heavy-ion collisions and hadron induced reactions, microscopic transport models are necessary. The properties of hadrons in nuclear medium are related to the issues of the interaction potential between hadron and nucleon, in-medium corrections of cross sections on hadron and resonance production, reabsorption process etc.

The properties of $\eta$ in nuclear medium has been investigated by several approaches, such as quark-meson coupling (QMC) model \cite{Ch91,Ts98}, chiral perturbation theory (ChPT) \cite{Wa97,Zh06,Ni08}, relativistic mean-field theory (RMF) \cite{Zh06}, chiral unitary approach etc \cite{In02}. Up to now, the strength of the $\eta$ optical potential at the normal nuclear density is not well refined, i.e., in the range from -20 MeV to -90 MeV predicted by different models \cite{Ni08}. Measurements on $\eta$ production in proton-nucleus collisions were performed in experiments \cite{Ch98,Ag13}. In this work, the dynamics of $\eta$ in proton and heavy-ion induced nuclear reactions is to be investigated within an isospin and momentum dependent transport model (Lanzhou quantum molecular dynamics transport model) \cite{Fe11,Fe12}. The impacts of the $\eta$ -nucleon potential and symmetry energy on the particle emission will be explored.

The article is organized as follows. In section II we give a brief description of the LQMD model. The in-medium effects on the $\eta$ production are discussed in section III. Conclusions are summarized in section IV.

\section{Model description}

It is well known that the wave function for each nucleon in the QMD-like models is represented by a Gaussian wave packet as follows \cite{Ai91}
\begin{equation}
\psi_i\left(\textbf{r,}t\right)=\frac{1}{\left(2\pi L\right)^{3/4}}\exp\left[-\frac{\left[\textbf{r}-\textbf{r}_i\left(t\right)\right]^2}{4L}\right]
\exp\left(\frac{i\textbf{p}_i\left(t\right)\cdot\textbf{r}}{\hbar}\right).
\end{equation}
Here $\textbf{r}_i\left(t\right)$ and $\textbf{p}_i\left(t\right)$ are the centers of the ith nucleon in the coordinate and momentum space, respectively. The $L$ is the square of the Gaussian wave packet width, which depends on the mass number of the nucleus being the form of $L=(0.92+0.08A^{1/3})^{2}$ fm$^{2}$ \cite{Fe08}. The total N-body wave function is assumed to be the direct product of the coherent states, where the antisymmetrization is neglected. After performing Wigner transformation for Eq. (1), we get the Wigner density as:
\begin{equation}
f\left(\textbf{r,p,}t\right)=\sum_i{f_i}\left(\textbf{r,p,}t\right)
\end{equation}
with
\begin{eqnarray}
f_i\left(\textbf{r,p,}t\right)=\frac{1}{\left(\pi\hbar\right)^3}\exp\left[-\frac{\left[\textbf{r}-\textbf{r}_i\left(t\right)\right]^2}{2L}-\frac{\left[\textbf{P}-\textbf{P}_i\left(t\right)\right]^2\cdot 2L}{\hbar^2}\right]
\nonumber \\
\end{eqnarray}
Then the density distributions in coordinate and momentum space are given by:
\begin{eqnarray}
\rho\left(\textbf{r},t\right) =\sum_i{\frac{1}{\left(2\pi L\right)^{3/2}} \exp\left[-\frac{\left[\textbf{r}-\textbf{r}_i\left(t\right)\right]^2}{2L}\right]}
\end{eqnarray}
\begin{eqnarray}
g\left(\textbf{p},t\right)=\sum_i{\left(\frac{2L}{\pi\hbar^2}\right)^{3/2}\exp\left[-\frac{\left[\textbf{p}-\textbf{p}_i\left(t\right)\right]^2\cdot 2L}{\hbar^2}\right]}
\end{eqnarray}
respectively, where the sum runs over all nucleons in the reaction systems.

In the LQMD model, the dynamics of the resonances ($\Delta$(1232), N*(1440), N*(1535)), hyperons ($\Lambda$, $\Sigma$, $\Xi$, $\Omega$) and mesons ($\pi$, $\eta$, $K$, $\overline{K}$) is described via hadron-hadron collisions, decays of resonances, mean-field potentials, and corrections on threshold energies of elementary cross sections \cite{Fe12,Fe15}. The temporal evolutions of the baryons (nucleons and resonances) and mesons in the reaction system under the self-consistently generated mean-field are governed by Hamilton's equations of motion, which read as
\begin{eqnarray}
\dot{\mathbf{p}}_{i} = -\frac{\partial H}{\partial\mathbf{r}_{i}}, \quad \dot{\mathbf{r}}_{i} = \frac{\partial H}{\partial\mathbf{p}_{i}}
\end{eqnarray}
The Hamiltonian of baryons consists of the relativistic energy, the effective interaction potential and the momentum dependent part as follows:
\begin{equation}
H_{B}=\sum_{i}\sqrt{\textbf{p}_{i}^{2}+m_{i}^{2}}+U_{int}+U_{mom}
\end{equation}
Here the $\textbf{p}_{i}$ and $m_{i}$ represent the momentum and the mass of the baryons.

The effective interaction potential is composed of the Coulomb interaction and the local interaction potential
\begin{equation}
U_{int}=U_{Coul}+U_{loc}
\end{equation}
The Coulomb interaction potential is written as
\begin{equation}
U_{Coul}=\frac{1}{2}\sum_{i,j,j\neq i} \frac{e_{i}e_{j}}{r_{ij}}erf(r_{ij}/\sqrt{4L})
\end{equation}
where the $e_{j}$ is the charged number including protons and charged resonances. The $r_{ij}=|\mathbf{r}_{i}-\mathbf{r}_{j}|$ is the relative distance of two charged particles.

The local interaction potential is derived from the Skyrme energy-density functional as the form of
$U_{loc}=\int V_{loc}(\rho(\mathbf{r}))d\mathbf{r}$. The energy-density functional reads
\begin{eqnarray}
V_{loc}(\rho) = && \frac{\alpha}{2}\frac{\rho^{2}}{\rho_{0}} +
\frac{\beta}{1+\gamma}\frac{\rho^{1+\gamma}}{\rho_{0}^{\gamma}} + E_{sym}^{loc}(\rho)\rho\delta^{2}
 + \frac{g_{sur}}{2\rho_{0}}(\nabla\rho)^{2}    \nonumber \\
 &&   +\frac{g_{sur}^{iso}}{2\rho_{0}}[\nabla(\rho_{n}-\rho_{p})]^{2}
\end{eqnarray}
where the $\rho_{n}$, $\rho_{p}$ and $\rho=\rho_{n}+\rho_{p}$ are the neutron, proton and total densities, respectively, and the $\delta=(\rho_{n}-\rho_{p})/(\rho_{n}+\rho_{p})$ being the isospin asymmetry. The coefficients $\alpha$, $\beta$, $\gamma$, $g_{sur}$, $g_{sur}^{iso}$ and $\rho_{0}$ are set to be the values of -215.7 MeV, 142.4 MeV, 1.322, 23 MeV fm$^{2}$, -2.7 MeV fm$^{2}$ and 0.16 fm$^{-3}$, respectively. A Skyrme-type momentum-dependent potential is used in the LQMD model \cite{Fe11}
\begin{eqnarray}\label{Vmom}
U_{mom}=&& \frac{1}{2\rho_{0}}\sum_{i,j,j\neq i}\sum_{\tau,\tau'}C_{\tau,\tau'}\delta_{\tau,\tau_{i}}\delta_{\tau',\tau_{j}}\int\int\int d \textbf{p}d\textbf{p}'d\textbf{r}   \nonumber \\
&& \times f_{i}(\textbf{r},\textbf{p},t) [\ln(\epsilon(\textbf{p}-\textbf{p}')^{2}+1)]^{2} f_{j}(\textbf{r},\textbf{p}',t)
\end{eqnarray}
Here $C_{\tau,\tau}=C_{mom}(1+x)$, $C_{\tau,\tau'}=C_{mom}(1-x)$ ($\tau\neq\tau'$) and the isospin symbols $\tau$($\tau'$) represent proton or neutron. The parameters $C_{mom}$ and $\epsilon$ were determined by fitting the real part of optical potential as a function of incident energy from the proton-nucleus elastic scattering data. In the calculation, we take the values of 1.76 MeV, 500 c$^{2}$/GeV$^{2}$ for the $C_{mom}$ and $\epsilon$, respectively, which result in the effective mass $m^{\ast}/m$=0.75 in nuclear medium at saturation density for symmetric nuclear matter. The parameter $x$ as the strength of the isospin splitting with the value of -0.65 is taken in this work, which leads to the mass splitting of $m^{\ast}_{n}>m^{\ast}_{p}$ in nuclear medium.  A compression modulus of K=230 MeV for isospin symmetric nuclear matter is concluded in the LQMD model.

The Hamiltonian of $\eta$ meson is constructed as follows:
\begin{eqnarray}
H_{\eta} =\sum_{i=1}^{N_{\eta}}{\sqrt{m_{\eta}^{2}+\textbf{p}_{i}^{2}}+V_{\eta}^{opt}\left(\textbf{p}_i,\rho_i\right)}
\end{eqnarray}
The optical potential $V_{\eta}^{opt}$ is
\begin{eqnarray}
V_{\eta}^{opt}= \sqrt{\left(m_{\eta}^{*}\right)^2+\textbf{p}_{i}^{2}}-\sqrt{m_{\eta}^{2}+\textbf{p}_{i}^{2}}.
\end{eqnarray}
The effective mass of $\eta$ meson in nuclear medium reads \cite{Zh06}
\begin{equation}
m_{\eta}^{*}=\sqrt{\left(m_{\eta}^{2}-\frac{\Sigma_{\eta N}}{f^2}\rho_s\right)/ \left(1+\frac{\kappa}{f^2}\rho_{s}\right)}.
\end{equation}
Here the pion decay constant $f_{\pi}=$92.4 MeV, $\Sigma_{\eta N}=$280 MeV, $\kappa =$0.4 fm, the $\eta$ mass $m_{\eta}$=547 MeV and $\rho_{s}$ being the scalar nucleon density. The value of $V_{\eta}^{opt}=$ -94 MeV is obtained with zero momentum and saturation density $\rho=\rho_{0}$. Shown in Fig. 1 is the optical potential as functions of momentum and baryon density, respectively. The strength of the potential increases with the baryon density, but decreases with the $\eta$ momentum. It is shown that the potential will influence the $\eta$ dynamics in dense nuclear matter.

\begin{figure}
\begin{center}
{\includegraphics*[width=1.\textwidth]{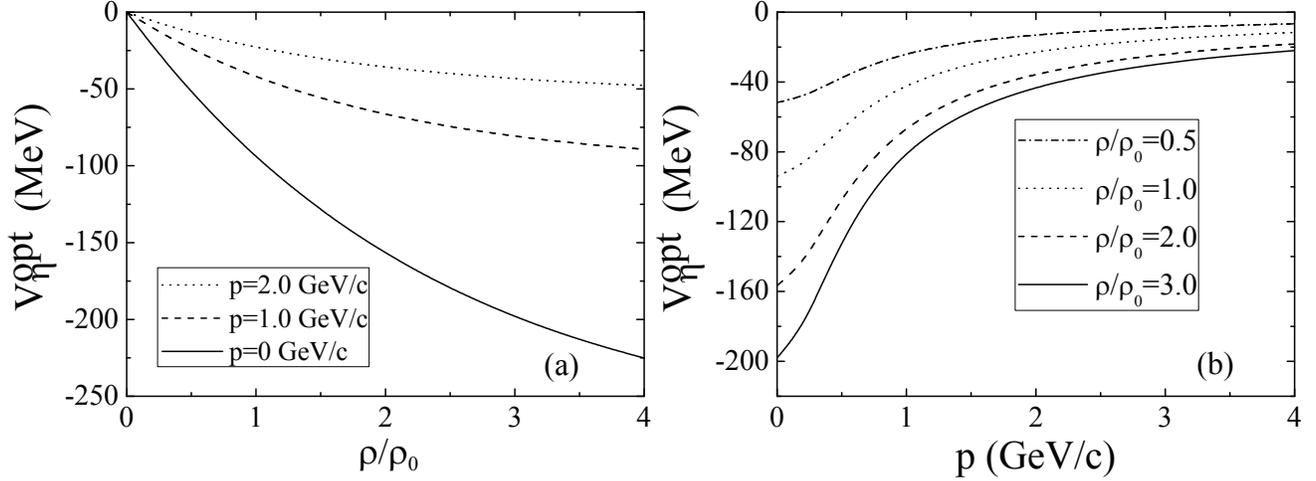}}
\end{center}
\caption{The density and momentum dependence of the $\eta$ optical potential. }
\end{figure}

The scattering in two-particle collisions is performed by using a Monte Carlo procedure, in which the probability to be a channel in a collision is calculated by its contribution of the channel cross section to the total cross section. The primary products in nucleon-nucleon (NN) collisions are the resonances of $\Delta$(1232), $N^{\ast}$(1440), and $N^{\ast}$(1535). We have included the reaction channels as follows:
\begin{eqnarray}
&& NN \leftrightarrow N\triangle, \quad  NN \leftrightarrow NN^{\ast}, \quad  NN
\leftrightarrow \triangle\triangle,  \Delta \leftrightarrow N\pi,          \nonumber \\
&&  N^{\ast} \leftrightarrow N\pi,   \quad  NN \leftrightarrow NN\pi (s-state),  \nonumber \\
&& N^{\ast}(1535) \leftrightarrow N\eta
\end{eqnarray}
The cross sections for $N^{\ast}$(1535) production can be estimated from the empirical $\eta$ yields \cite{Gy93}, such as $\sigma(pp(nn) \rightarrow NN^{\ast}(1535)) \approx 2\sigma(pp(nn) \rightarrow pp(nn)\eta)=$ (a) $0.34 s_{r}/(0.253+s_{r}^{2})$, (b) $0.4 s_{r}/(0.552+s_{r}^{2})$ and (c) $0.204 s_{r}/(0.058+s_{r}^{2})$ in mb and $s_{r}=\sqrt{s}-\sqrt{s_{0}}$ with $\sqrt{s}$ being the invariant energy in GeV and $\sqrt{s_{0}}=2m_{N}+m_{\eta}=2.424$ GeV \cite{Fe15}. The case (b) is used in this work. The $np$ cross sections are about 3 times larger than that for $nn$ or $np$. We have taken a constant width of $\Gamma$=150 MeV for the $N^{\ast}$(1535) decay, and $N^{\ast}$(1535) has half probabilities decay into the $\eta$ meson and half to $\pi$.

\section{Results and discussion}

At the considered energies in this work, the $\eta$ meson is produced from the decay of N$^{\ast}$(1535). Therefore, the properties of N$^{\ast}$(1535) in nuclear medium are dominant on the $\eta$ dynamics. We managed the mean-field potential for N$^{\ast}$(1535) similar to the ones of nucleons. The $\pi^{0}$ and $\eta$ have similar properties in nuclear medium. Both of the mesons are neutral particles and decay from $N^{\ast}$(1535) with the same probability. In this work, we did not include the isospin, density and momentum dependent pion-nucleon potential in Ref. \cite{Fe15}. The $\eta/\pi^{0}$ ratios in heavy-ion collisions are calculated as a test of our approach. The values in light ($^{12}$C+$^{12}$C), intermediate-mass ($^{40}$Ar+$^{40}$Ca, $^{86}$Kr+$^{90}$Zr) and heavy symmetric ($^{197}$Au+$^{197}$Au) systems are compared with the available data from the Two-Arm Photon Spectrometer (TAPS) collaboration \cite{Av03} as shown in Table I. Basically, the calculations are consistent with the available data. The ratios increase with participating numbers of colliding partners and incident energies.

\begin{table}
\caption{Comparison of the $\eta/\pi^{0}$ ratios of calculations and the available data \cite{Av03} in heavy-ion collisions} \vspace*{-10pt}
\begin{center}
\def\temptablewidth{1.\textwidth}
{\rule{\temptablewidth}{1pt}}
\begin{tabular*}{\temptablewidth}{@{\extracolsep{\fill}}ccccccc}
&E$_{beam}$(\emph{A} GeV)   & reaction systems  & experimental data ($\%$)  & calculated values ($\%$)   \\
\hline
&1.0 & C+C & 0.57$\pm$0.14 & 1.186  \\
&1.0 & Ar+Ca & 1.3$\pm$0.8 & 1.321   \\
&1.0 & Kr+Zr & 1.3$\pm$0.6 & 1.663   \\
&1.0 & Au+Au & 1.4$\pm$0.6 & 2.087  \\
&1.5 & Ar+Ca & 2.2$\pm$0.4 & 2.823  \\
\end{tabular*}
{\rule{\temptablewidth}{1pt}}
\end{center}
\end{table}

The extraction of the in-medium properties of $\eta$ in proton induced reactions has advantage in comparison to heavy-ion collisions, in which particles are produced around the saturation densities (0.8$\rho_{0}\sim$1.2$\rho_{0}$) \cite{Fe14}. The in-medium properties of $\eta$ are related to the issues of resonance production in nuclear medium, i.e., N$^{\ast}$(1535), resonance-nucleon and $\eta$-nucleon potentials. Here, we concentrate on the $\eta$-nucleon interaction from proton-nucleus collisions and its impact on $\eta$ production  near threshold energies ($E_{th}(\eta)$=1.26 GeV). Shown in Fig. 2 is the rapidity distributions in collisions of protons on $^{12}$C and $^{40}$Ca at the kinetic energy of 1.5 GeV.  It should be noticed that the $\eta$-nucleon potential enhances the backward $\eta$ emissions. The transverse momentum spectra are calculated as shown in Fig. 3. The high-momentum yields are reduced because of the enhancement of reabsorption process in $\eta$-nucleon scattering with the eta potential. Similar structure is also found from the kinetic-energy spectra of invariant cross sections as shown in Fig. 4. The effect becomes more pronounced with increasing the mass numbers of target nuclei because of larger collision probabilities between $\eta$ and nucleons. The invariant spectra become steeper with the potential for both cases, which show a lower local temperature and longer interaction time of $\eta$ meson in nuclear medium after inclusion of the $\eta$-nucleon potential. The strength of the $\eta$-nucleon potential is not well refined up to now, which is of significance in the formation of possible $\eta$-bound state ($\eta$-nucleus). The results would be helpful for extracting the $\eta$-nucleon potential from proton-nucleus collisions in the near future experiments.

\begin{figure}
\begin{center}
{\includegraphics*[width=1.\textwidth]{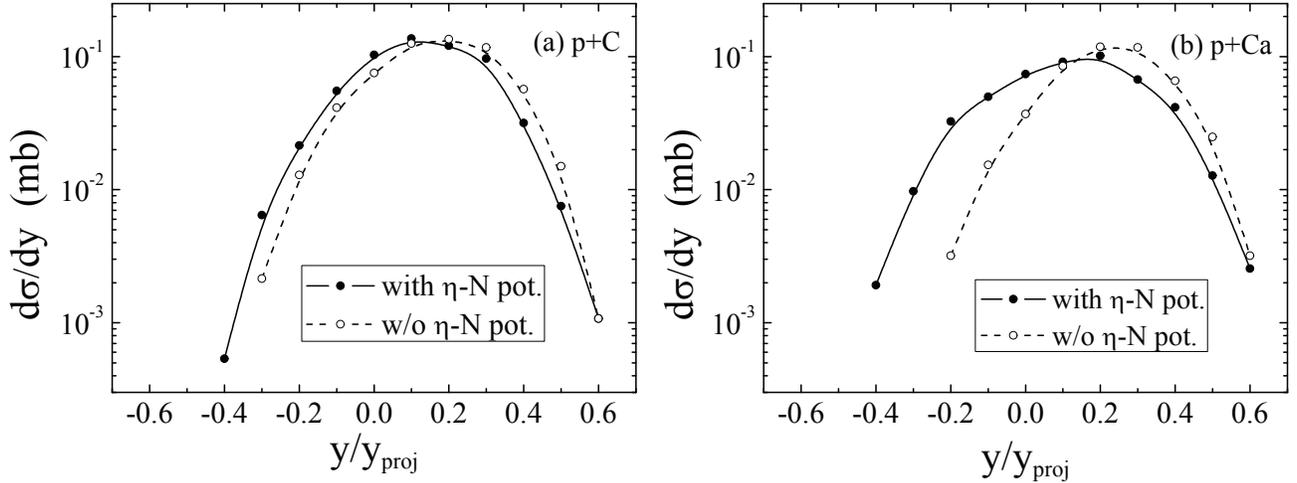}}
\end{center}
\caption{Rapidity distributions in proton induced reactions at incident energy of 1.5 GeV. The solid and dashed lines are shown for guiding eyes.}
\end{figure}

\begin{figure}
\begin{center}
{\includegraphics*[width=1.\textwidth]{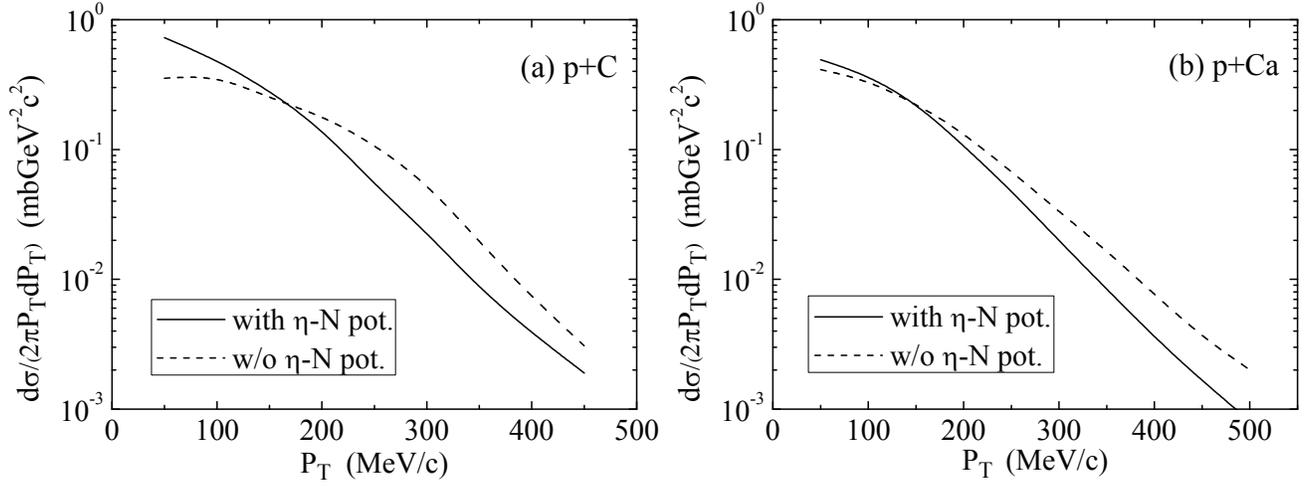}}
\end{center}
\caption{Transverse momentum distributions in proton induced reactions at incident energy of 1.5 GeV.}
\end{figure}

\begin{figure}
\begin{center}
{\includegraphics*[width=1.\textwidth]{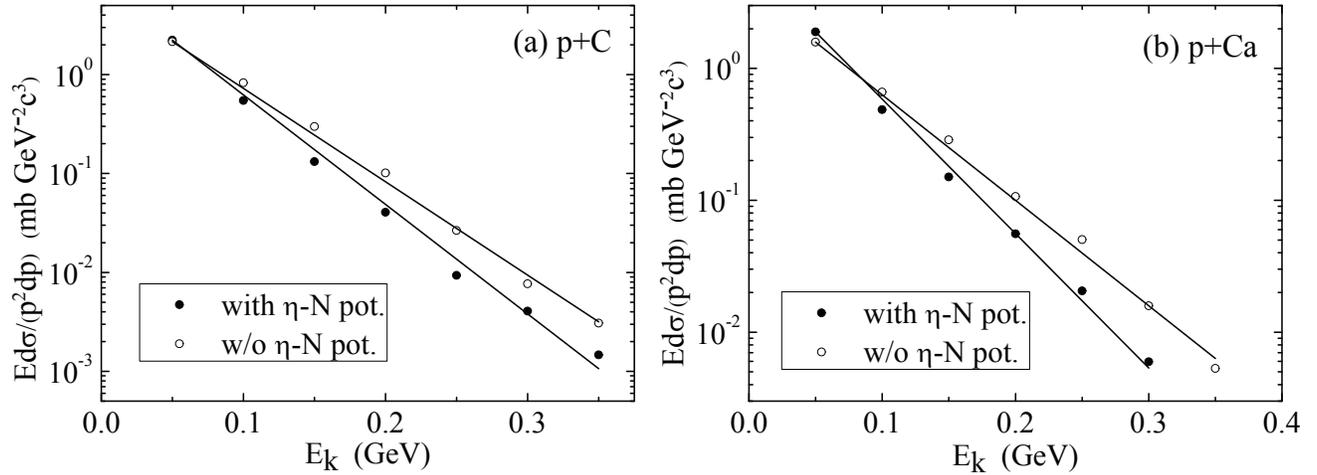}}
\end{center}
\caption{Kinetic energy spectra of invariant cross sections in collisions of protons on $^{12}$C and $^{40}$Ca at incident energy of 1.5 GeV, respectively. The solid lines are shown for guiding eyes.}
\end{figure}

\section{Conclusions}

The dynamics of $\eta$ meson produced in proton-induced nuclear reactions has been investigated within the LQMD transport model. The $\eta$ mesons are produced via the decay of N$^{\ast}(1535)$. The reabsorption of $\eta$ and N$^{\ast}(1535)$ by surrounding nucleons dominates the $\eta$ distributions in phase space. The attractive $\eta$-nucleon potential enhances the reabsorption process in nuclear medium, which leads to the reduction of forward emissions. The yields of $\eta$ in collisions of protons on nuclei at high-momenta (kinetic energies) are reduced with the $\eta$-nucleon potential. The effect becomes more pronounced with increasing the mass numbers of target nuclei.

\end{document}